Stability and Motion around Equilibrium Points in the Rotating Plane-Symmetric Potential Field

Yu Jiang[1, 2], Hexi Baoyin[2], Xianyu Wang[2], Hengnian Li[1]
1. School of Aerospace Engineering, Tsinghua University, Beijing 100084, China
2. State Key Laboratory of Astronautic Dynamics, Xi'an Satellite Control Center, Xi'an 710043, China
Y. Jiang (✉) e-mail: jiangyu_xian_china@163.com (corresponding author)

**Abstract.** This study presents a study of equilibrium points, periodic orbits, stabilities, and manifolds in a rotating plane-symmetric potential field. It has been found that the dynamical behaviour near equilibrium points is completely determined by the structure of the submanifolds and subspaces. The non-degenerate equilibrium points are classified into twelve cases. The necessary and sufficient conditions for linearly stable, non-resonant unstable and resonant equilibrium points are established. Furthermore, the results show that a resonant equilibrium point is a Hopf bifurcation point. In addition, if the rotating speed changes, two non-degenerate equilibria may collide and annihilate each other. The theory developed here is lastly applied to two particular cases, motions around a rotating, homogeneous cube and the asteroid 1620 Geographos. We found that the mutual annihilation of equilibrium points occurs as the rotating speed increases, and then the first surface shedding begins near the intersection point of the –x axis and the surface. The results can be applied to planetary science, including the birth and evolution of the minor bodies in the Solar system, the rotational breakup and surface mass shedding of asteroids, etc.

## 1 Introduction

Space missions to minor bodies [1-5], such as asteroids, comets, and satellites around planets in the solar system, as well as the discovery of binary asteroids, make the dynamical behaviour in the vicinity of non-spherically shaped bodies (such as a massive inhomogeneous straight segment) a subject of increasing interest [6-7]. Some space missions consider flying a spacecraft around an asteroid and even landing on its surface [8], leading importance to the study of the dynamics in the potential field of an asteroid. In addition, the dynamics of a large-mass-ratio binary asteroid [9] that can be modelled as a massless particle flying around a large and irregularly shaped body (such as Ida and Dactyl, [10]) is also relevant to research concerning the motion near an irregularly shaped body.

The classical method of modelling celestial bodies is to expand the gravity potential using the Legendre polynomial series [11]; this method can provide a good approximation to nearly spherically shaped celestial bodies when the series is sufficiently long [12]. However, many minor bodies, such as asteroids, comets, and satellites around planets, have irregular shapes. For space missions to minor bodies, it is necessary to calculate the gravitational field of these irregular-shaped bodies. However, the method of the Legendre polynomial series does not converge at certain points [13-14] or regions [15]. Several methods are used to eliminate this difficulty.

Werner [16] developed a method that uses a polyhedron to model irregularly shaped bodies such as asteroids, comet nuclei, and small planetary satellites and then applied this method to calculate the gravitational field of the inner Martian satellite Phobos. Subsequently, the polyhedron method was applied to several asteroids, including asteroids 4769 Castalia [17], 4179 Toutatis [18], 216 Kleopatra [19-25] and the binary near-Earth asteroid (66391) 1999 KW4 [26-28].

However, the polyhedron model contains many free parameters and is highly complex; some simply shaped models may also yield good approximations for some bodies [13]. That is, although the polyhedron model offers higher precision for quantitatively analysing and computing the dynamical behaviour in the vicinity of some asteroids, the qualitative analysis of the dynamical behaviour in the vicinity of certain asteroids still may be achieved by considering simply shaped bodies. Thus, Elipe and Lara [13] have used a finite straight segment to study the equilibria, periodic-orbit families, bifurcations and stability regions in phase space in the vicinity of asteroid 433 Eros. Broucke and Elipe [29] have discussed the potential, periodic-orbit families and bifurcations in the potential field of a solid circular ring. Blesa [14] has presented several families of periodic orbits in the plane of a triangular plate and a square plate. Alberti and Vidal [30] have calculated the potential of a homogeneous annulus disk and have studied the orbital motion near the disk. Fukushima [31] has derived the acceleration of a uniform ring or disk. Liu et al. [32] have investigated the locations and linear stability of equilibria, periodic orbits around equilibria and heteroclinic orbits in the gravitational field of a rotating homogeneous cube. Li et al. [33] have investigated the locations and linear stability of equilibrium points as well as periodic orbits around equilibrium points in the vicinity of a rotating dumbbell-shaped body. These simply shaped bodies and potential fields, including the logarithmic gravity field [34], the straight segment [6][8][13][35], the solid circular ring [29][36], the triangular plate and the square plate [14], the homogeneous annulus disk [30-31], the homogeneous cube [32, 37-39], the dumbbell-shaped body [33], the classical rotating dipole model [40-42], and the dipole segment model [43] are all plane-symmetric. The relative equilibria of spacecrafts in the second degree and order-gravity field [44-45] are different from the equilibria in the above studies. The second degree and order-gravity field is also a rotating plane-symmetric potential field. The relative equilibria of spacecrafts include the equilibria of the position and the attitude, and the position is also the relative equilibria in the plane-symmetric potential field.

In this work, we are interested in the study of the dynamics of orbits in a rotating plane-symmetric gravitational field (unless explicitly stated otherwise, all discussions concern the dynamics in this type of gravitational field), including equilibrium points, periodic orbits, and manifolds. The dynamical behaviours in the xy plane and the z axis are decoupled for the plane-symmetric case, and the topological classifications for the plane-symmetric case are different with the general cases. Thus we focus on the plane-symmetric case in the current study.

The linearised equations of motion relative to an equilibrium point are derived and investigated in Sect. 2. Furthermore, the characteristic equation of equilibrium points is presented. In Sect. 3, the structure of the submanifolds and subspaces near an equilibrium point are studied, which fixes the motion state near the equilibrium point. It is found that there are twelve cases for the non-degenerate equilibrium points in the plane-symmetric potential field of a rotating plane-symmetric body. The necessary and sufficient conditions for linearly stable, non-resonant unstable and resonant equilibrium points are presented. If the rotating speed varies, two non-degenerate equilibria may collide and change to one degenerate equilibrium point and then annihilate each other.

The theory developed in this study is then applied to the motion in the gravitational potential of a rotating homogeneous cube and asteroid 1620 Geographos [42] in Sect. 4. In the gravitational potential of a rotating homogeneous cube, it is found that there are two families of periodic orbits on the xy plane near equilibrium points E1, E3, E5 and E7, and there is only one family of periodic orbits on the xy plane near the equilibrium points E2, E4, E6 and E8. While the rotation speed of asteroid 1620 Geographos varies, the number of equilibrium points will change from five to three to one. The positions of mutual annihilation of equilibrium points are inside the body of asteroid 1620 Geographos. The results can be applied to the scientific research of the birth and evolution of the Solar System and its minor bodies [46-47], the rotational breakup of asteroids and comets [48], as well as the surface mass shedding of asteroids [49] in the future Human mission to asteroids [48][50].

## 2 Equations of motion

### 2.1 Equations of motion in the arbitrary body-fixed frame

The potential field of a rotating plane-symmetric body satisfies

$$U\left(x,y,z\right)=U\left(x,y,-z\right), \tag{1}$$

where $\left(x,y,z\right)$ is the coordinates in the body-fixed coordinate system, and $U$ is the potential of the body.

Consider the motion of a massless particle in the potential field of a rotating plane-symmetric body; the dynamical system is a Hamiltonian system.

The equations of motion of the particle relative to the body can be written as

$$\ddot{\mathbf{r}}+2\boldsymbol{\omega}\times\dot{\mathbf{r}}+\boldsymbol{\omega}\times(\boldsymbol{\omega}\times\mathbf{r})+\dot{\boldsymbol{\omega}}\times\mathbf{r}+\frac{\partial U(\mathbf{r})}{\partial\mathbf{r}}=0, \tag{2}$$

where $\mathbf{r}$ is the body-fixed vector from the centre of mass of the body to the particle, $\boldsymbol{\omega}$ is the rotational-angular-velocity vector of the body relative to the inertial frame of reference, and $\frac{\partial U(\mathbf{r})}{\partial\mathbf{r}}$ is the gradient of the potential. If $\boldsymbol{\omega}=0$, then the body is fixed and has no rotation.

The Jacobian integral $H$ is defined as

$$H=\frac{1}{2}\dot{\mathbf{r}}\cdot\dot{\mathbf{r}}-\frac{1}{2}(\boldsymbol{\omega}\times\mathbf{r})(\boldsymbol{\omega}\times\mathbf{r})+U(\mathbf{r}). \tag{3}$$

$H$ is time invariant if and only if $\boldsymbol{\omega}$ is time invariant. When $H$ is time invariant, it is also called the Jacobian constant.

The effective potential can be defined as [20][51]

$$V(\mathbf{r})=-\frac{1}{2}(\boldsymbol{\omega}\times\mathbf{r})(\boldsymbol{\omega}\times\mathbf{r})+U(\mathbf{r}), \tag{4}$$

which satisfies

$$V(x,y,z)=V(x,y,-z). \tag{5}$$

For a uniformly rotating body, equation (2) can be simplified to

$$\ddot{\mathbf{r}}+2\boldsymbol{\omega}\times\dot{\mathbf{r}}+\frac{\partial V(\mathbf{r})}{\partial\mathbf{r}}=0. \tag{6}$$

The body is assumed to be uniformly rotating throughout this paper. The body-fixed frame can be defined via a set of orthonormal right-handed unit vectors $\mathbf{e}$:

$$\mathbf{e} \equiv \begin{Bmatrix} \mathbf{e}_{\mathrm{x}} \\ \mathbf{e}_{\mathrm{y}} \\ \mathbf{e}_{\mathrm{z}} \end{Bmatrix}. \tag{7}$$

The frame of reference that is used throughout this study is the body-fixed frame. Let $\omega$ be the modulus of the vector $\boldsymbol{\omega}$; in addition, consider that the vector $\boldsymbol{\omega}$ can be written as $\boldsymbol{\omega} = \omega_{\mathrm{x}}\mathbf{e}_{\mathrm{x}} + \omega_{\mathrm{y}}\mathbf{e}_{\mathrm{y}} + \omega_{\mathrm{z}}\mathbf{e}_{\mathrm{z}}$. The equilibrium points are the critical points of the effective potential $V(\mathbf{r})$.

The linearised equations of motion relative to the equilibrium point can be written as

$$\begin{aligned} &\ddot{\xi} + 2\omega_y\dot{\zeta} - 2\omega_z\dot{\eta} + V_{\mathrm{xx}}\xi + V_{\mathrm{xy}}\eta = 0 \\ &\ddot{\eta} + 2\omega_z\dot{\xi} - 2\omega_x\dot{\zeta} + V_{\mathrm{yx}}\xi + V_{\mathrm{yy}}\eta = 0 \\ &\ddot{\zeta} + 2\omega_x\dot{\eta} - 2\omega_y\dot{\xi} + V_{\mathrm{zz}}\zeta = 0 \end{aligned} . \tag{8}$$

Where $\xi = x - x_{\mathrm{E}},\ \eta = y - y_{\mathrm{E}}, \zeta = z - z_{\mathrm{E}}$. Here $\left(x_{\mathrm{E}}, y_{\mathrm{E}}, z_{\mathrm{E}}\right)$ represents the location of the equilibrium point.

The characteristic equation follows:

$$\begin{vmatrix} \lambda^2 + V_{\mathrm{xx}} & -2\omega_z\lambda + V_{\mathrm{xy}} & 2\omega_y\lambda \\ 2\omega_z\lambda + V_{\mathrm{yx}} & \lambda^2 + V_{\mathrm{yy}} & -2\omega_x\lambda \\ -2\omega_y\lambda & 2\omega_x\lambda & \lambda^2 + V_{\mathrm{zz}} \end{vmatrix} = 0. \tag{9}$$

Furthermore, it can be rewritten as a sextic equation in $\lambda$:

$$\begin{aligned} &\lambda^6 + \left(V_{\mathrm{xx}} + V_{\mathrm{yy}} + V_{\mathrm{zz}} + 4\omega_x^2 + 4\omega_y^2 + 4\omega_z^2\right)\lambda^4 + \\ &\left(V_{\mathrm{xx}}V_{\mathrm{yy}} + V_{\mathrm{yy}}V_{\mathrm{zz}} + V_{\mathrm{zz}}V_{\mathrm{xx}} - V_{xy}^2 + 8\omega_x\omega_y V_{xy} + 4\omega_x^2 V_{xx} + 4\omega_y^2 V_{yy} + 4\omega_z^2 V_{zz}\right)\lambda^2 \\ &+ \left(V_{\mathrm{xx}}V_{\mathrm{yy}}V_{\mathrm{zz}} - V_{\mathrm{zz}}V_{xy}^2\right) = 0 \end{aligned} , \tag{10}$$

where $\lambda$ denotes the eigenvalues of Eq. (8). The linear stability of the equilibrium point is determined by the six roots of Eq. (10). Let $\lambda_i \left( i = 1, 2, \cdots, 6 \right)$ represent the roots of Eq. (10).

**2.2 General equations of motion in the special body-fixed frame**

If the axis of rotation $\boldsymbol{\omega}$ and the axis of coordinates $\mathbf{e}_z$ are coincident, then the body-fixed coordinate system is defined by $\boldsymbol{\omega} = \omega \mathbf{e}_z$, and the linearised Eq. (8) can be simplified to

$$\begin{aligned} &\ddot{\xi} - 2\omega\dot{\eta} + V_{xx}\xi + V_{xy}\eta = 0 \\ &\ddot{\eta} + 2\omega\dot{\xi} + V_{yx}\xi + V_{yy}\eta = 0 , \\ &\ddot{\zeta} + V_{zz}\zeta = 0 \end{aligned} \tag{11}$$

while Eq. (9) can be simplified to

$$\begin{vmatrix} \lambda^2 + V_{xx} & -2\omega\lambda + V_{xy} & 0 \\ 2\omega\lambda + V_{yx} & \lambda^2 + V_{yy} & 0 \\ 0 & 0 & \lambda^2 + V_{zz} \end{vmatrix} = 0 \tag{12}$$

or

$$\left( \lambda^2 + V_{zz} \right) \left[ \lambda^4 + \left( V_{xx} + V_{yy} + 4\omega^2 \right) \lambda^2 + V_{xx}V_{yy} - V_{xy}^2 \right] = 0 . \tag{13}$$

The equation $\lambda^2 + V_{zz} = 0$ determines the eigenvalues on the z-axis, while the equation $\lambda^4 + \left( V_{xx} + V_{yy} + 4\omega^2 \right) \lambda^2 + \left( V_{xx}V_{yy} - V_{xy}^2 \right) = 0$ determines the eigenvalues in the xy plane. The effective potential reduces to $V\left(\mathbf{r}\right) = U\left(\mathbf{r}\right) - \frac{\omega^2}{2}\left( x^2 + y^2 \right)$, so $\lim\limits_{|\mathbf{r}| \to +\infty} \left[ V\left(\mathbf{r}\right) + \frac{\omega^2}{2}\left( x^2 + y^2 \right) \right] = 0$ ; besides, $V\left(\mathbf{r}\right) = C$ denotes a 2-dimensional curved surface, where $C$ is a constant. The asymptotic surface of $V = V\left(\mathbf{r}\right)$ is a circular cylindrical surface that can be expressed as $V^{*} = -\frac{\omega^2}{2}\left(x^2 + y^2\right)$, where the radius of the circular cylindrical surface has the form $\frac{\sqrt{2}}{2}\omega\sqrt{x^2 + y^2}$. In addition, the Jacobian integral then takes the form $H = U + \frac{1}{2}\left(\dot{x}^2 + \dot{y}^2 + \dot{z}^2\right) - \frac{\omega^2}{2}\left(x^2 + y^2\right)$, and the Lagrange function takes the form $L = \frac{1}{2}\left(\dot{x}^2 + \dot{y}^2 + \dot{z}^2\right) + \frac{1}{2}\omega^2\left(x^2 + y^2\right) + \omega\left(x\dot{y} - \dot{x}y\right) - U$. If $\omega$ is time invariant, then $H$ is a constant along the solutions, meaning that the integral of the relative energy is conserved.

## 3. Periodic orbits and submanifolds near equilibrium points

### 3.1 Eigenvalues and structure of submanifolds

If the Hessian matrix of the effective potential at the equilibrium point has full rank, then the equilibrium point is called a *non-degenerate equilibrium point*. Based on the discussion above, one can obtain the following theorem regarding the topological classification of the non-degenerate equilibrium points in the plane-symmetric potential field of a rotating plane-symmetric body that the rotation is uniform and the axis of rotation is perpendicular to the plane of symmetry.

**Theorem 1.** There are twelve cases for the non-degenerate equilibrium points in the plane-symmetric potential field of a rotating plane-symmetric body. The eigenvalues of the linearised equations of motion relative to a non-degenerate equilibrium point in the potential field of a rotating plane-symmetric body can be classified topologically on the complex plane as shown in Fig. 1.

The magnitude of eigenvalues is arbitrary for each topographical classification, for instance, Case e and h in XY plane, four eigenvalues $\sigma + i\tau$, $\sigma - i\tau$, $-\sigma + i\tau$, $-\sigma - i\tau$, where $\sigma, \tau \in \mathrm{R}; \sigma, \tau > 0$, can be located anywhere.

Fig. 1 shows the topological classifications of six eigenvalues on the complex plane. The abscissa axis is the x axis, and the ordinate axis is the y axis. The coordinate planes labelled XY show the four eigenvalues in the set $C_{XY}$, while the coordinate planes labelled Z show the two eigenvalues in the set $C_Z$.

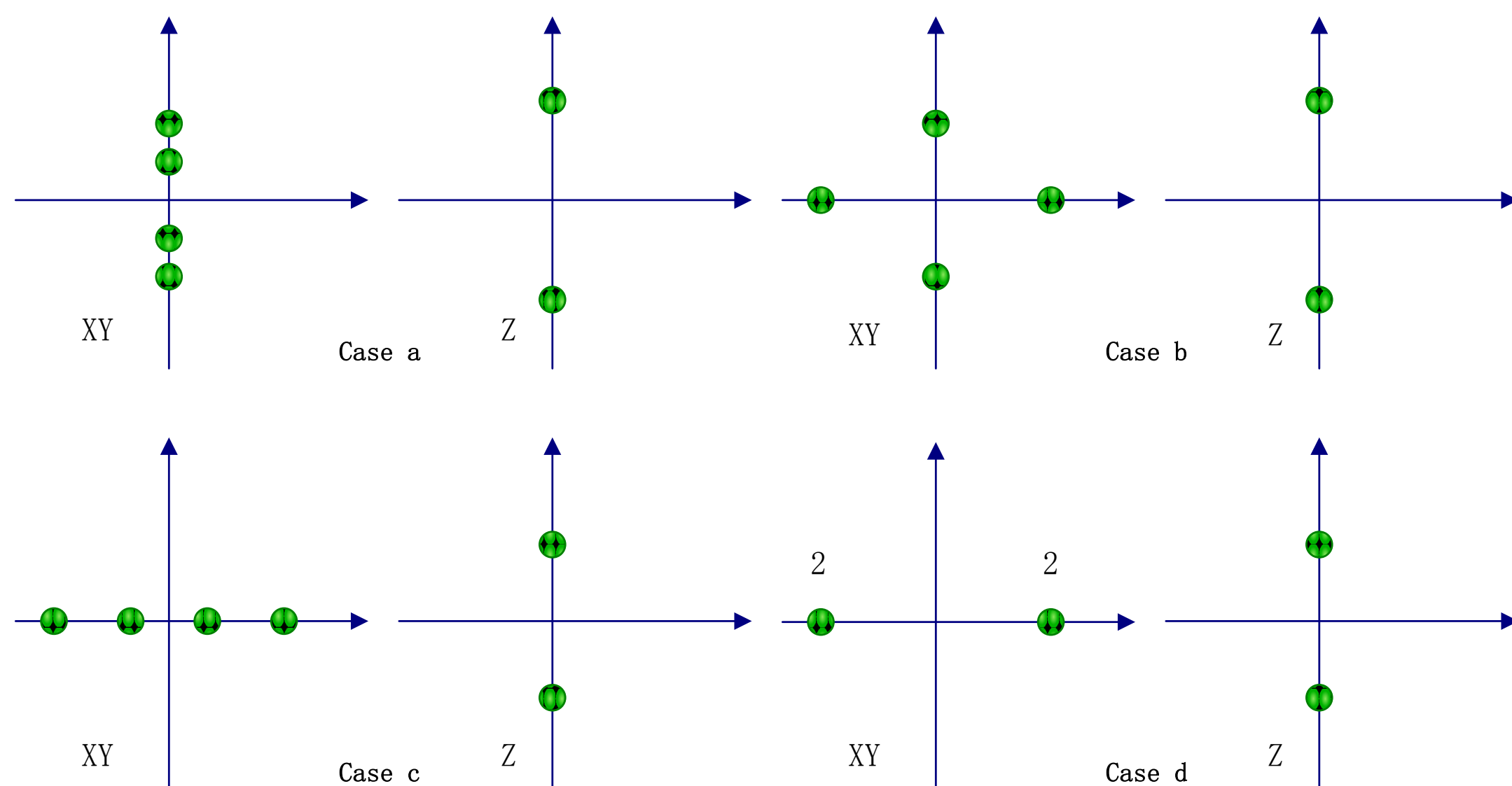

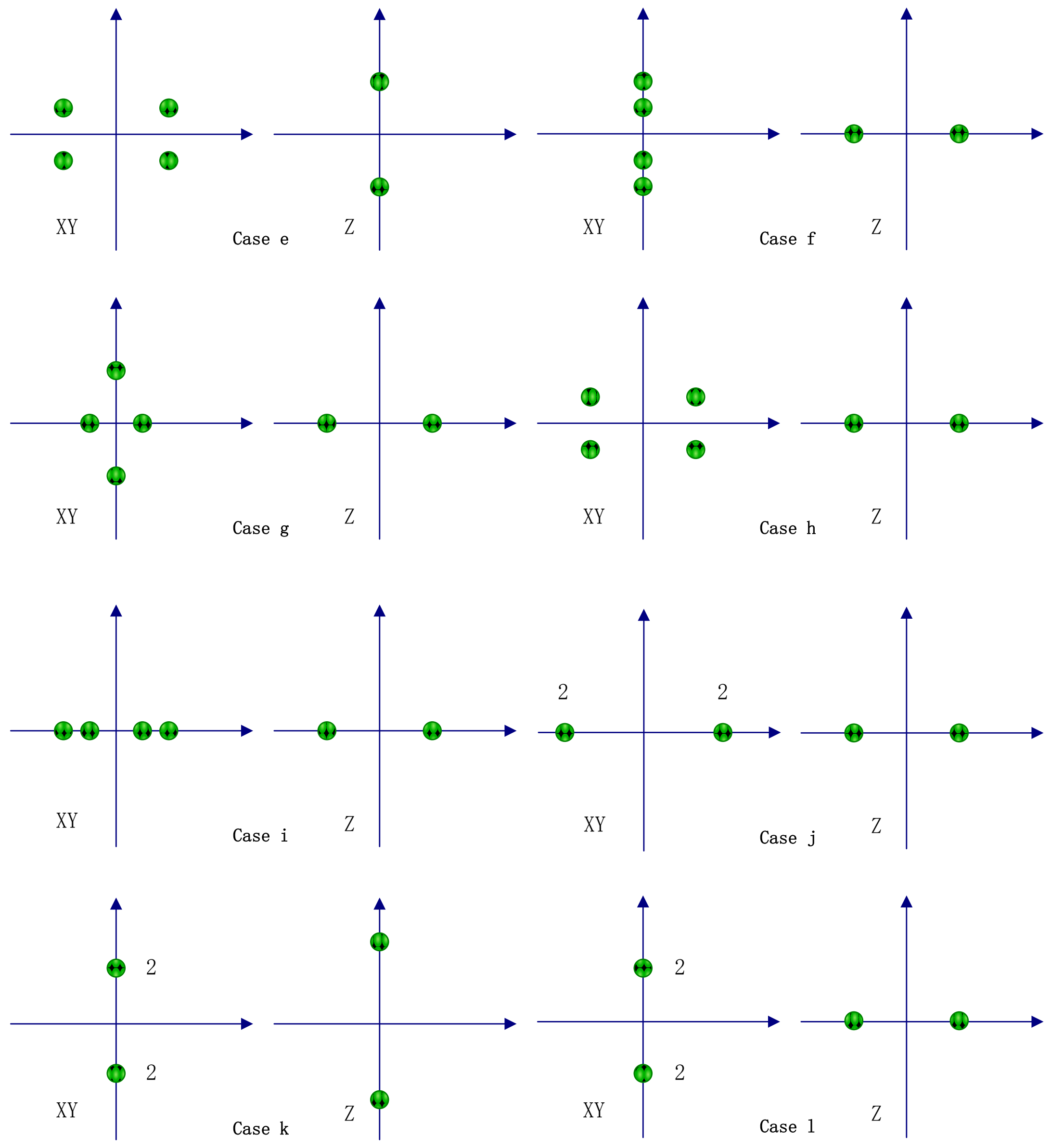

Fig. 1 The topological classification of six eigenvalues on the complex plane

Theorem 1 describes the structure of the submanifold and the stable and unstable behaviours of a particle near the equilibrium points. Only Case a leads to linearly stable equilibrium points, and Cases b-j lead to non-resonant unstable equilibrium points. For Cases k and l, because the resonant manifold and the resonant subspace exist, the equilibrium point is resonant. Considering that the structures of the submanifold and the subspace are fixed by the characteristics of the equilibrium points, it can be concluded that equilibrium points with resonant manifolds are resonant equilibrium points. For Cases a-e and Case k, the motion component that lies along the z axis near the equilibrium point is linearly stable. For Cases f-j and Case l, the motion component along the z axis is unstable. For Case a and Case f, the motion component in the xy plane is linearly stable. For Case d and Case j, the motion component in the xy plane is unstable, and the phase diagrams of the motion in the xy plane correspond to motion near a real saddle. Thus, one can obtain the following:

**Corollary 1.** In the potential field of a rotating plane-symmetric body, an equilibrium point is linearly stable if and only if it belongs to Case a, an equilibrium point is unstable and non-resonant if and only if it belongs to one of the Cases b-j, and an equilibrium point is resonant if and only if it belongs to one of the Cases k or l.

For an equilibrium point of the Hamiltonian, if there exist 2 pairs of imaginary eigenvalues $\pm\lambda_j, \pm\lambda_k$ such that $\frac{\lambda_j}{\lambda_k} = \frac{p}{q}$, where $p$ and $q$ are positive integers, then the equilibrium point is a resonant equilibrium point [52]. For the resonant equilibrium point, we have the following Corollary:

**Corollary 2.** In the potential field of a rotating plane-symmetric body, the motion component along the z axis of a particle near an equilibrium point is linearly stable if and only if the equilibrium point belongs to one of the Cases a-e or to Case k; the motion component in the xy plane of a particle near an equilibrium point is linearly stable if and only if the equilibrium point belongs to either Case a or Case f.

Corollary 1 presents the necessary and sufficient conditions for linearly stable, non-resonant unstable and resonant equilibrium points. In addition, the motion component along the z axis of a particle near a linearly stable equilibrium point exhibits simple harmonic motion if and only if the equilibrium point belongs to one of the Cases a-e or to Case k.

**3.2 Motion near non-resonant equilibrium points**

Non-resonant equilibrium points include linearly stable equilibrium points and non-resonant unstable equilibrium points. Corollary 2 states that linearly stable equilibrium points arise only in Case a. In this section, more properties of Case a are discussed.

In Case a, there are three pairs of imaginary eigenvalues of the equilibrium point, which is linearly stable. The motion of a particle relative to the equilibrium point follows a quasi-periodic orbit, and the motion component along the z axis of a particle near such a linearly stable equilibrium point corresponds to simple harmonic motion.

There are three families of periodic orbits, each of which has a period of $T_1 = \frac{2\pi}{\beta_1}, T_2 = \frac{2\pi}{\beta_2}$ or $T_3 = \frac{2\pi}{\beta_3}$. There are four families of quasi-periodic orbits on the $k$-dimensional tori $T^k \left(k = 2,3\right)$, which can be expressed as follows:

$$\begin{cases} \xi = C_{\xi 1} \cos \beta_1 t + S_{\xi 1} \sin \beta_1 t + C_{\xi 2} \cos \beta_2 t + S_{\xi 2} \sin \beta_2 t \\ \eta = C_{\eta 1} \cos \beta_1 t + S_{\eta 1} \sin \beta_1 t + C_{\eta 2} \cos \beta_2 t + S_{\eta 2} \sin \beta_2 t \\ \zeta = 0 \end{cases},$$

$$\begin{cases} \xi = C_{\xi 1} \cos \beta_1 t + S_{\xi 1} \sin \beta_1 t \\ \eta = C_{\eta 1} \cos \beta_1 t + S_{\eta 1} \sin \beta_1 t \\ \zeta = C_{\zeta 3} \cos \beta_3 t + S_{\zeta 3} \sin \beta_3 t \end{cases},$$

$$\begin{cases} \xi = C_{\xi 2} \cos \beta_2 t + S_{\xi 2} \sin \beta_2 t \\ \eta = C_{\eta 2} \cos \beta_2 t + S_{\eta 2} \sin \beta_2 t \\ \zeta = C_{\zeta 3} \cos \beta_3 t + S_{\zeta 3} \sin \beta_3 t \end{cases} \quad \text{and}$$

$$\begin{cases} \xi = C_{\xi 1} \cos \beta_1 t + S_{\xi 1} \sin \beta_1 t + C_{\xi 2} \cos \beta_2 t + S_{\xi 2} \sin \beta_2 t \\ \eta = C_{\eta 1} \cos \beta_1 t + S_{\eta 1} \sin \beta_1 t + C_{\eta 2} \cos \beta_2 t + S_{\eta 2} \sin \beta_2 t \\ \zeta = C_{\zeta 3} \cos \beta_3 t + S_{\zeta 3} \sin \beta_3 t \end{cases}.$$

Non-resonant unstable equilibrium points can be classified into nine cases, which are Cases b-j.

### 3.3 Resonant equilibrium points

The resonant equilibrium point is a Hopf bifurcation point when the parameter varies [53], and the Hopf bifurcation can transfer to chaos[54], in other words, the resonant equilibrium point can lead to the chaotic motion near the resonant equilibrium point. The appearance and disappearance of periodic-orbit families are found near resonant equilibrium points with parametric variation. Define $\boldsymbol{\mu} = \boldsymbol{\mu}(t)$ as a parameter of the effective potential that is dependent on the time $t$, which means that

$$V(\boldsymbol{\mu}, \mathbf{r}) = -\frac{1}{2}(\boldsymbol{\omega}(\boldsymbol{\mu}) \times \mathbf{r})(\boldsymbol{\omega}(\boldsymbol{\mu}) \times \mathbf{r}) + U(\boldsymbol{\mu}, \mathbf{r}). \quad (14)$$

If $\boldsymbol{\mu} = \boldsymbol{\mu}(t)$ changes, the rotational-angular-velocity vector of the body relative to the inertial frame, which is denoted as $\boldsymbol{\omega}$, also changes, then the effective potential $V(\boldsymbol{\mu}, \mathbf{r})$ is time-variant relative to the vector $\boldsymbol{\omega}$. Define $\boldsymbol{\mu}_0 = \boldsymbol{\mu}(t_0)$; let the open neighbourhood of $\boldsymbol{\mu}_0$ be $G_N(\boldsymbol{\mu}_0)$. The periodic orbits in the xy plane near a resonant equilibrium point are dense. Theorem 3 indicates that the system around a resonant equilibrium point with parametric variation is sensitive to initial conditions and topological mixing. Thus, the dynamical system around such a resonant equilibrium point with parametric variation is chaotic. If a dynamical system is chaotic, then it is sensitive to initial conditions and topologically mixing, in addition, it has dense periodic orbits. The motion around the non-resonant unstable equilibrium point is unstable. This is the difference between chaotic motion and the motion for the non-resonant unstable equilibrium.

The periodic orbits are dense means that in an arbitrarily small neighbourhood of the point in the phase space, there exist a point represents a periodic orbit" [55-56]. Consider the orbits in the xy plane and periodic orbits. The motion in the xy plane near a resonant equilibrium point can be expressed as follows:

$$\begin{cases} \xi = C_{\xi 1}\cos\beta_1 t + S_{\xi 1}\sin\beta_1 t + P_{\xi 1} t\cos\beta_1 t + Q_{\xi 1} t\sin\beta_1 t \\ \eta = C_{\eta 1}\cos\beta_1 t + S_{\eta 1}\sin\beta_1 t + P_{\eta 1} t\cos\beta_1 t + Q_{\eta 1} t\sin\beta_1 t \\ \dot{\xi} = -C_{\xi 1}\beta_1\sin\beta_1 t + S_{\xi 1}\beta_1\cos\beta_1 t + P_{\xi 1}\cos\beta_1 t + Q_{\xi 1}\sin\beta_1 t - P_{\xi 1} t\beta_1\sin\beta_1 t + Q_{\xi 1} t\beta_1\cos\beta_1 t \\ \dot{\eta} = -C_{\eta 1}\beta_1\sin\beta_1 t + S_{\eta 1}\beta_1\cos\beta_1 t + P_{\eta 1}\cos\beta_1 t + Q_{\eta 1}\sin\beta_1 t - P_{\eta 1} t\beta_1\sin\beta_1 t + Q_{\eta 1} t\beta_1\cos\beta_1 t \end{cases} . \quad (15)$$

Consider the periodic orbit

$$\begin{cases} \xi = \widehat{C}_{\xi 1} \cos \beta_1 t + \widehat{S}_{\xi 1} \sin \beta_1 t \\ \eta = \widehat{C}_{\eta 1} \cos \beta_1 t + \widehat{S}_{\eta 1} \sin \beta_1 t \\ \dot{\xi} = -\widehat{C}_{\xi 1} \beta_1 \sin \beta_1 t + \widehat{S}_{\xi 1} \beta_1 \cos \beta_1 t \\ \dot{\eta} = -\widehat{C}_{\eta 1} \beta_1 \sin \beta_1 t + \widehat{S}_{\eta 1} \beta_1 \cos \beta_1 t \end{cases}, \tag{16}$$

where the coefficients satisfy

$$\begin{cases} \widehat{C}_{\xi 1} = C_{\xi 1} \\ \widehat{C}_{\eta 1} = C_{\eta 1} \\ \widehat{S}_{\xi 1} \beta_1 = S_{\xi 1} \beta_1 + P_{\xi 1} \\ \widehat{S}_{\eta 1} \beta_1 = S_{\eta 1} \beta_1 + P_{\eta 1} \end{cases}. \tag{17}$$

Then, the distance between these two orbits in 4-dimensional phase space is 0. The above proofs are only for the linearised motions. This leading to the following conclusion: the linearised periodic motions in the xy plane near the resonant equilibrium point are dense.

Considering a non-degenerate resonant equilibrium point, we have the following:

**Theorem 2.** If the non-degenerate equilibrium point $\mathbf{r}_0 = \boldsymbol{\tau}(\boldsymbol{\mu}_0) = \boldsymbol{\tau}_0$ in a rotating plane-symmetric potential field is resonant, then it is a branching point in the presence of persistently acting parameter variation. Furthermore, Hopf bifurcation occurs at the non-degenerate resonant equilibrium point.

**Proof**: The eigenvalues of a non-degenerate equilibrium point have only the forms $\pm\alpha\left(\alpha \in \mathrm{R}, \alpha > 0\right)$, $\pm i\beta\left(\beta \in \mathrm{R}, \beta > 0\right)$, and $\pm\sigma \pm i\tau\left(\sigma, \tau \in \mathrm{R}; \sigma, \tau > 0\right)$. Therefore, the pure imaginary eigenvalues on the xy plane will not leave the imaginary axis before colliding, which means that Case k may perhaps go to Case a or Case e, while Case l may perhaps go to Case f or Case h. Thus, equilibrium points that belong to Case k or Case l are Hopf bifurcation points. □

For the movement of the eigenvalues, we have the following theorem.

**Theorem 3.** For a non-degenerate equilibrium point in a rotating plane-symmetric potential field, the following statements hold:

a) The eigenvalues in Case a can only move to Case k.

b) The eigenvalues in Case k can only move to Case a or Case e.

c) The eigenvalues in Case f can only move to Case l.

d) The eigenvalues in Case l can only move to Case f or Case h.

e) If the non-degenerate equilibrium point belongs to Case a, when the parameter $\boldsymbol{\mu}$ changes, the movement of the eigenvalues on the xy plane may follow the pattern $\text{Case a} \rightarrow \text{Case k} \rightarrow \text{Case e}$.

f) If the non-degenerate equilibrium point belongs to Case e, when the parameter $\boldsymbol{\mu}$ changes, the movement of the eigenvalues on the xy plane may follow the pattern $\text{Case e} \rightarrow \text{Case k} \rightarrow \text{Case a}$ or $\text{Case e} \rightarrow \text{Case d} \rightarrow \text{Case c}$.

g) If the non-degenerate equilibrium point belongs to Case f, when the parameter $\boldsymbol{\mu}$ changes, the movement of the eigenvalues on the xy plane may follow the pattern $\text{Case f} \rightarrow \text{Case l} \rightarrow \text{Case h}$.

h) If the non-degenerate equilibrium point belongs to Case h, when the parameter $\boldsymbol{\mu}$ changes, the movement of the eigenvalues on the xy plane may follow the pattern $\text{Case h} \rightarrow \text{Case l} \rightarrow \text{Case f}$ or

$\text{Case h} \rightarrow \text{Case j} \rightarrow \text{Case i}$.

Consider the parameter $\boldsymbol{\mu} = \boldsymbol{\mu}(t)$ varies which correspond to the pattern $\text{Case a} \rightarrow \text{Case k} \rightarrow \text{Case e}$, the equilibrium point which belongs to $\text{Case k}$ is resonant, that means an arbitrarily small change of the initial trajectory may lead to significantly different future behavior. One can get the same result for Case l.

We also have the following:

**Comment 1.** For a non-degenerate equilibrium point in a rotating plane-symmetric potential field, if the non-degenerate equilibrium point belongs to Case c, then when the parameter $\boldsymbol{\mu}$ changes, the movement of the eigenvalues may follow the pattern $\text{Case c} \rightarrow \text{Case d} \rightarrow \text{Case e}$; if the non-degenerate equilibrium point belongs to Case i, when the parameter $\boldsymbol{\mu}$ changes, the movement of the eigenvalues may follow the pattern $\text{Case i} \rightarrow \text{Case j} \rightarrow \text{Case h}$.

Fig. 2a shows the movement of eigenvalues on the xy plane that corresponds to $\text{Case a} \rightarrow \text{Case k} \rightarrow \text{Case e}$, while Fig. 2b shows the movement of eigenvalues on the xy plane that corresponds to $\text{Case e} \rightarrow \text{Case k} \rightarrow \text{Case a}$. When the movement of the eigenvalues follows the pattern $\text{Case a} \rightarrow \text{Case k} \rightarrow \text{Case e}$, the disappearance of periodic-orbit families is observed, i.e., the number of periodic-orbit families near the equilibrium point changes as follows: $3 \rightarrow 2 \rightarrow 1$. Conversely, when the movement of the eigenvalues follows the pattern $\text{Case e} \rightarrow \text{Case k} \rightarrow \text{Case a}$, the appearance of periodic-orbit families is observed, i.e., the number of periodic-orbit families near the equilibrium point changes as follows: $1 \to 2 \to 3$.

Fig. 3a shows the movement of eigenvalues on the xy plane that corresponds to $\text{Case f} \to \text{Case l} \to \text{Case h}$, while Fig. 3b shows the movement of eigenvalues on the xy plane that corresponds to $\text{Case h} \to \text{Case l} \to \text{Case f}$. When the movement of the eigenvalues follows the pattern $\text{Case f} \to \text{Case l} \to \text{Case h}$, the disappearance of periodic-orbit families is observed, i.e., the number of periodic-orbit families near the equilibrium point changes as follows: $2 \to 1 \to 0$. Conversely, when the movement of the eigenvalues follows the pattern $\text{Case h} \to \text{Case l} \to \text{Case f}$, the appearance of periodic-orbit families is observed, i.e., the number of periodic-orbit families near the equilibrium point changes as follows: $0 \to 1 \to 2$.

Notice that the equilibrium points that belong to Case a are linearly stable, and the equilibrium points that belong to Case e are non-resonant and unstable; thus, we have the following corollary:

**Corollary 3.** In a rotating plane-symmetric potential field, if the non-degenerate equilibrium point $\mathbf{r}_0 = \boldsymbol{\tau}(\boldsymbol{\mu}_0) = \boldsymbol{\tau}_0$ belongs to Case k or Case l, then for any sufficiently small open neighbourhood $G_N(\boldsymbol{\mu}_0)$ of $\boldsymbol{\mu}_0$, $\exists \boldsymbol{\mu}_1, \boldsymbol{\mu}_2 \in G_N(\boldsymbol{\mu}_0)$, such that the equilibrium point $\boldsymbol{\tau}(\boldsymbol{\mu}_1)$ is linearly stable, and the equilibrium point $\boldsymbol{\tau}(\boldsymbol{\mu}_2)$ is non-resonant and unstable. That is, there is a function $\mathbf{r} = \boldsymbol{\tau}(\boldsymbol{\mu})$ such that $\boldsymbol{\tau}(\boldsymbol{\mu}_0) = \boldsymbol{\tau}_0$ and $\mathbf{F}(\boldsymbol{\mu}, \boldsymbol{\tau}) = 0$ for any $\boldsymbol{\mu} \in G_N(\boldsymbol{\mu}_0)$, and $\boldsymbol{\tau}(\boldsymbol{\mu})$ is an equilibrium point for any $\boldsymbol{\mu} \in G_N(\boldsymbol{\mu}_0)$. In addition, $\boldsymbol{\tau}(\boldsymbol{\mu}_1)$ is linearly stable, while $\boldsymbol{\tau}\left(\boldsymbol{\mu}_2\right)$ is non-resonant and unstable.

The neighborhood $G_N\left(\boldsymbol{\mu}_0\right)$ can be time-variant or time-invariant, the only condition for the neighborhood $G_N\left(\boldsymbol{\mu}_0\right)$ is that it is a sufficiently small open neighbourhood. The Corollary 3 investigates the local dynamical behaviours for the Case k and Case l. The following Section 3.4.1 and 3.4.2 give the structure of the submanifold and orbits near the equilibrium point.

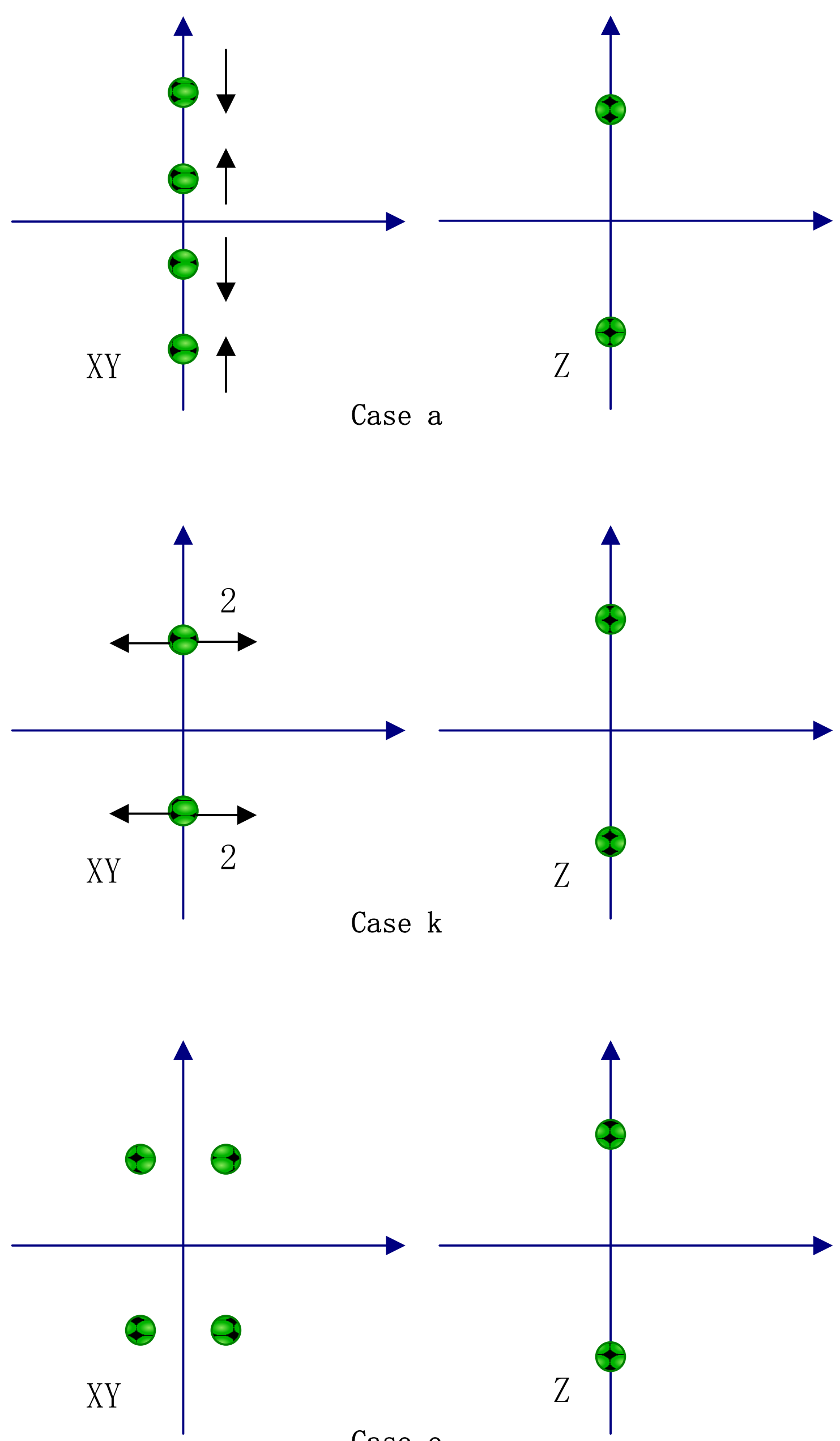

Fig. 2a The movement of eigenvalues on the xy plane that corresponds to

Case a $\rightarrow$ Case k $\rightarrow$ Case e

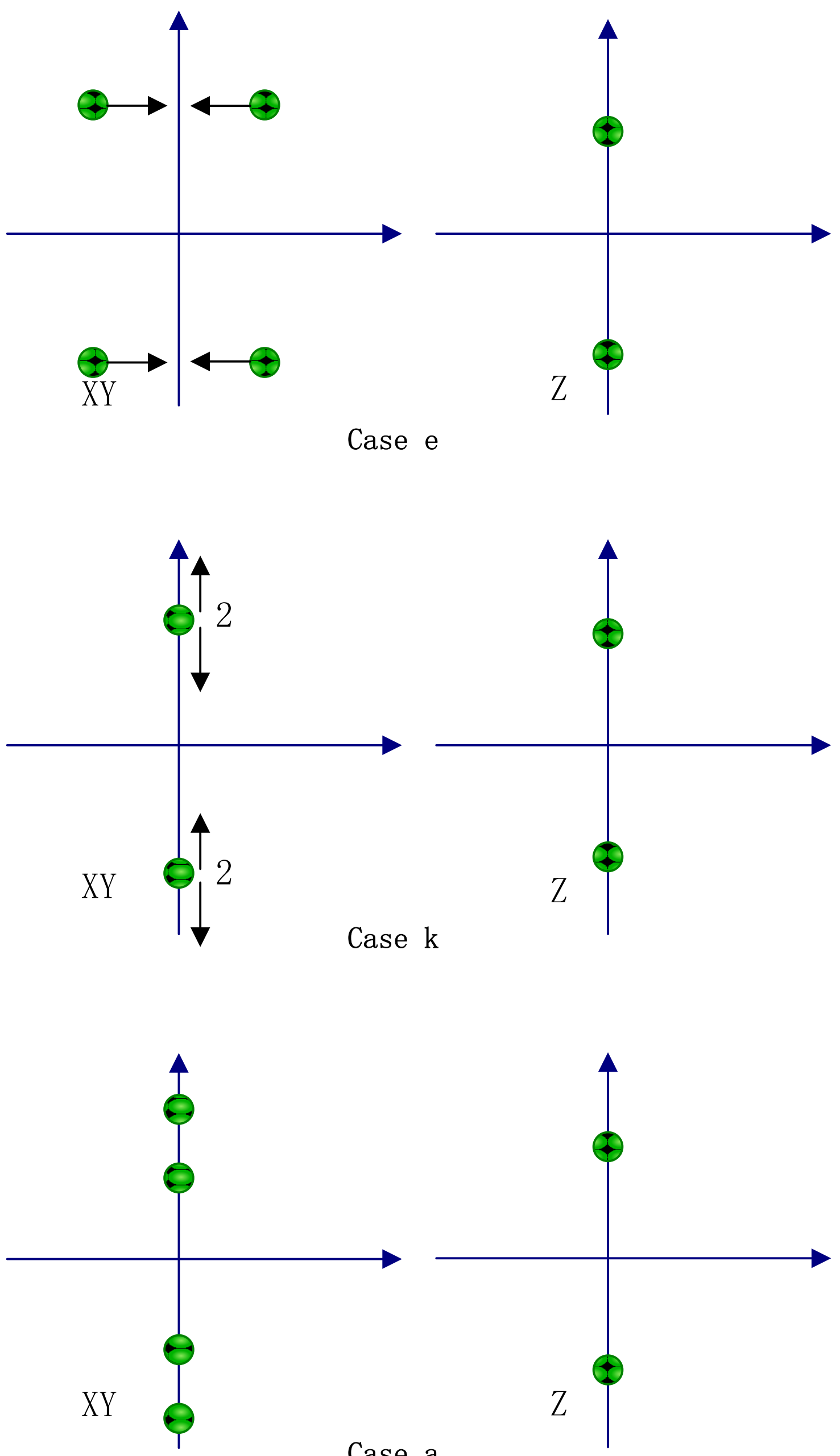

Fig. 2b The movement of eigenvalues on the xy plane that corresponds to

Case e $\rightarrow$ Case k $\rightarrow$ Case a

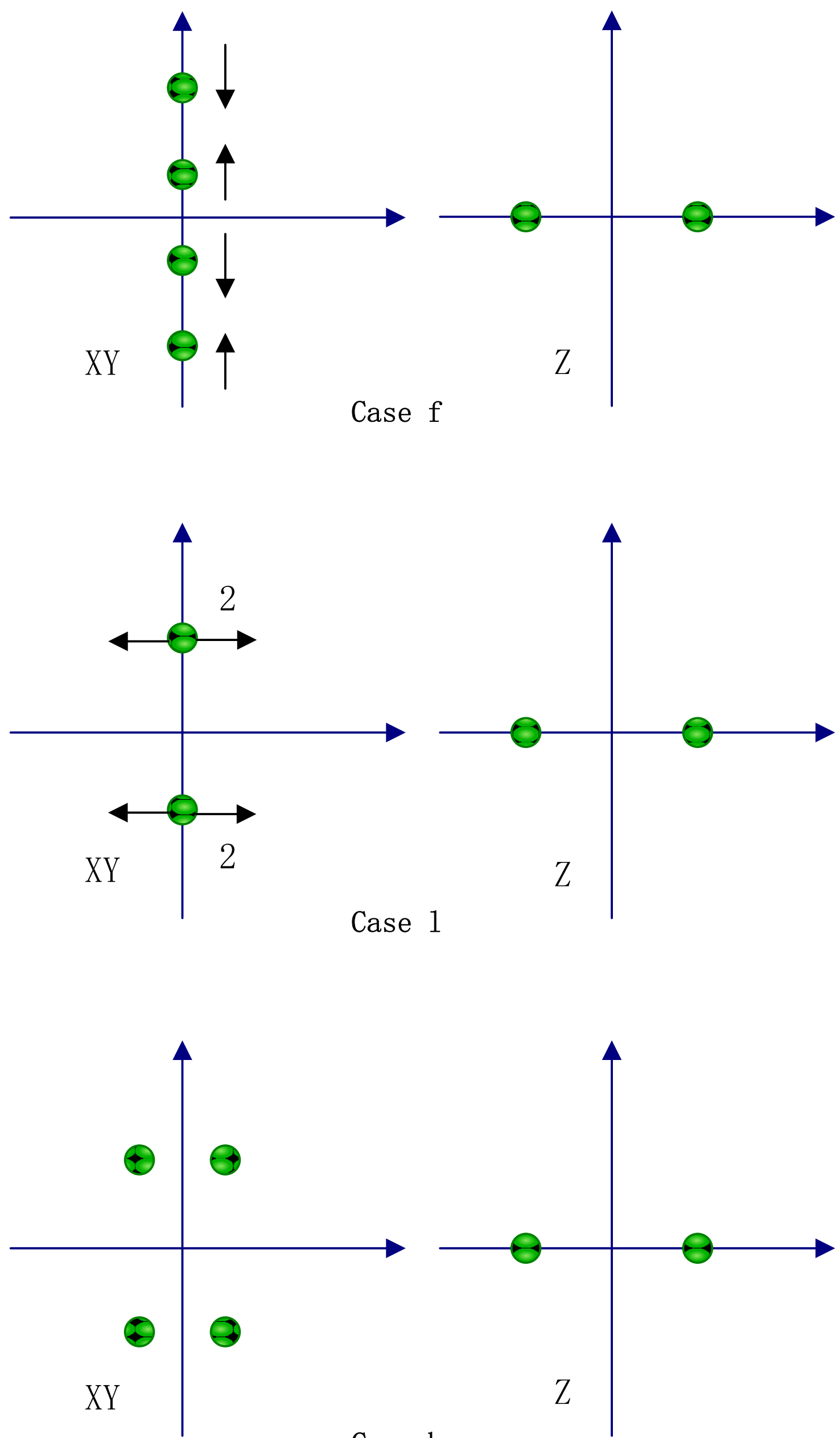

Fig. 3a The movement of eigenvalues on the xy plane that corresponds to

Case f $\rightarrow$ Case l $\rightarrow$ Case h

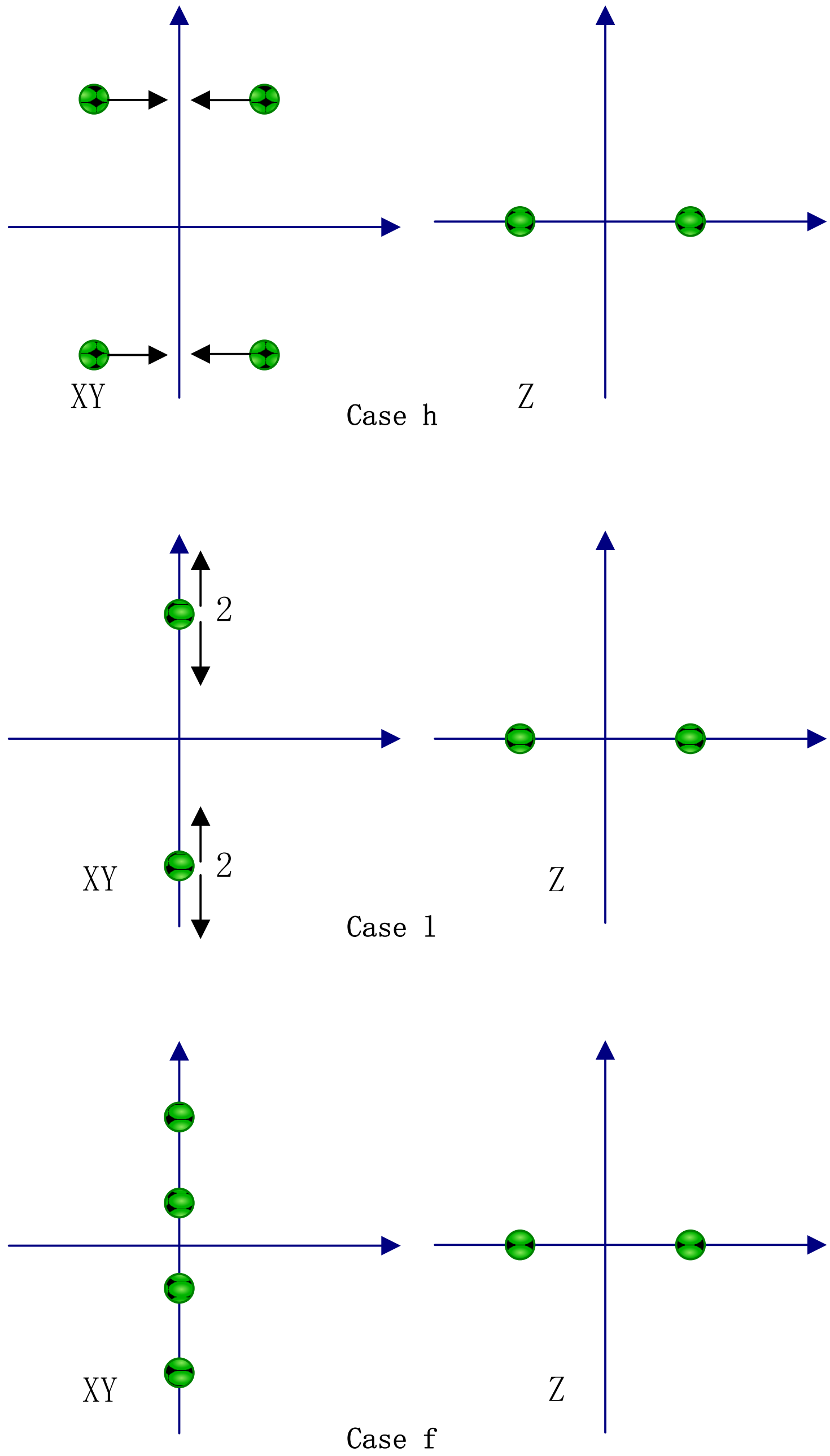

Fig. 3b The movement of eigenvalues on the xy plane that corresponds to

Case h $\rightarrow$ Case l $\rightarrow$ Case f

### 3.3.1 Case k

In this case, the quasi-periodic orbit near the equilibrium point can be expressed

as follows:

$$\begin{cases} \xi = C_{\xi 1}\cos\beta_1 t + S_{\xi 1}\sin\beta_1 t \\ \eta = C_{\eta 1}\cos\beta_1 t + S_{\eta 1}\sin\beta_1 t \\ \zeta = C_{\zeta 3}\cos\beta_3 t + S_{\zeta 3}\sin\beta_3 t \end{cases} . \tag{18}$$

There are two families of periodic orbits near the equilibrium point, which have the limiting periods $T_1 = \dfrac{2\pi}{\beta_1}$ and $T_3 = \dfrac{2\pi}{\beta_3}$ as the periodic orbits approach to the equilibrium point. One family of periodic orbits is in the xy plane, and the other is on the z axis.

**3.3.2 Case l**

In this case, the periodic orbit near the equilibrium point can be expressed as follows:

$$\begin{cases} \xi = C_{\xi 1}\cos\beta_1 t + S_{\xi 1}\sin\beta_1 t \\ \eta = C_{\eta 1}\cos\beta_1 t + S_{\eta 1}\sin\beta_1 t \\ \zeta = 0 \end{cases} . \tag{19}$$

There is only one family of periodic orbits near the equilibrium point, which have the limiting period $T_1 = \dfrac{2\pi}{\beta_1}$ as the periodic orbits approach the equilibrium point. The family of periodic orbits is in the xy plane, and there is no periodic orbit on the z axis.

**3.5 Mutual Annihilation of Equilibrium Points with a Variable Rotating Speed**

The rotating speed is the norm of the rotational-angular-velocity vector of the body. Let the rotating speed vary; consider the motion of equilibrium points. Eq.

(11) can be expressed by

$$\dot{\Psi} = \mathbf{g}(\Psi) = \begin{pmatrix} \mathbf{0}_{3\times3} & \mathbf{I}_{3\times3} \\ -\nabla^2 V & -\mathbf{G} \end{pmatrix} \begin{bmatrix} \boldsymbol{\mu} \\ \boldsymbol{\delta} \end{bmatrix} = \mathbf{P}\Psi , \tag{20}$$

Where $\boldsymbol{\delta}$ is the position of the particle relative to the equilibrium point, $\dot{\boldsymbol{\delta}} = \boldsymbol{\mu}$, and

$$\Psi = \begin{bmatrix} \boldsymbol{\mu} \\ \boldsymbol{\delta} \end{bmatrix}, \tag{21}$$

$$\mathbf{G} = \begin{pmatrix} 0 & -2\omega & 0 \\ 2\omega & 0 & 0 \\ 0 & 0 & 0 \end{pmatrix}, \tag{22}$$

$$\mathbf{P} = \begin{pmatrix} \mathbf{0}_{3\times3} & \mathbf{I}_{3\times3} \\ -\nabla^2 V & -\mathbf{G} \end{pmatrix}. \tag{23}$$

Denote $f(\mathbf{r}) = \nabla V$, then $\dfrac{df}{d\mathbf{r}} = \nabla^2 V$ and $\det \mathbf{P} = \det(\nabla^2 V)$.

Denote $\mathbf{S}_e$ as the open set, for the function $f(\mathbf{r})$, use the topological degree theory yields

$$\sum_{i=1}^{N} \left[ \operatorname{sgn} \prod_{j=1}^{6} \lambda_i (E_k) \right] = \deg\left(f, \quad \Xi, \quad (0, \quad 0, \quad 0)\right) = const , \tag{24}$$

where $N$ is the number of relative equilibria, $E_k$ is the $k$ th equilibrium point, $\lambda_j(E_k)$ is the $j$ th eigenvalue of $E_k$.

If an equilibrium point has zero eigenvalues, it is called the degenerate equilibrium point. Eq. (24) implies that the number of non-degenerate equilibria in the rotating plane-symmetric potential field varies in pairs. If the rotating speed varies, two non-degenerate equilibria may collide and change to one degenerate equilibrium point and then disappear. In other words, these two non-degenerate equilibria collide and change to one degenerate equilibrium point before mutual annihilation. The mutual annihilation of triple non-degenerate equilibria is impossible. If the equilibrium point $E_k$ is degenerate, $\operatorname{sgn}\prod_{j=1}^{6}\lambda_i\left(E_k\right)=0$. If the equilibrium point $E_k$ is non-degenerate, $\operatorname{sgn}\prod_{j=1}^{6}\lambda_i\left(E_k\right)=\pm 1$. When the non-degenerate equilibrium point $E_k$ belongs to case a or case e, $\operatorname{sgn}\prod_{j=1}^{6}\lambda_i\left(E_k\right)=1$. When the non-degenerate equilibrium point $E_k$ belongs to case f, $\operatorname{sgn}\prod_{j=1}^{6}\lambda_i\left(E_k\right)=-1$. Only a non-degenerate equilibrium point with $\operatorname{sgn}\prod_{j=1}^{6}\lambda_i\left(E_k\right)=1$ and another non-degenerate equilibrium point with $\operatorname{sgn}\prod_{j=1}^{6}\lambda_i\left(E_k\right)=-1$ can annihilate each other. For example, two non-degenerate equilibria belong to case e and case f collide and change to two degenerate equilibrium points, and then annihilate.

## 4. Applications

In this section, the theory developed in the previous sections is applied to the motion in the gravitational potential of a rotating homogeneous cube and asteroid 1620 Geographos. The potential field of the cube is plane-symmetric and the potential field of asteroid 1620 Geographos is approximating plane-symmetric.

### 4.1 Application to the rotating homogeneous cube

The motion around a rotating homogeneous cube is a particular case of the motion of a particle in the potential field of a rotating plane-symmetric body, and the theory of motion near a rotating homogeneous cube is a particular corollary of the theory developed in the previous sections. Fig. 4 shows the positions of the equilibrium points in the body-fixed frame of the rotating homogeneous cube when the ratio of gravitational acceleration to centrifugal acceleration is equal to 1. On the x axis and y axis, the eigenvalues of the equilibrium points E1, E3, E5, and E7 [33] are $\pm 0.697i$ and $\pm 0.789i$; and the eigenvalues of the equilibrium points E2, E4, E6, and E8 are $\pm 0.545i$ and $\pm 1.187$.

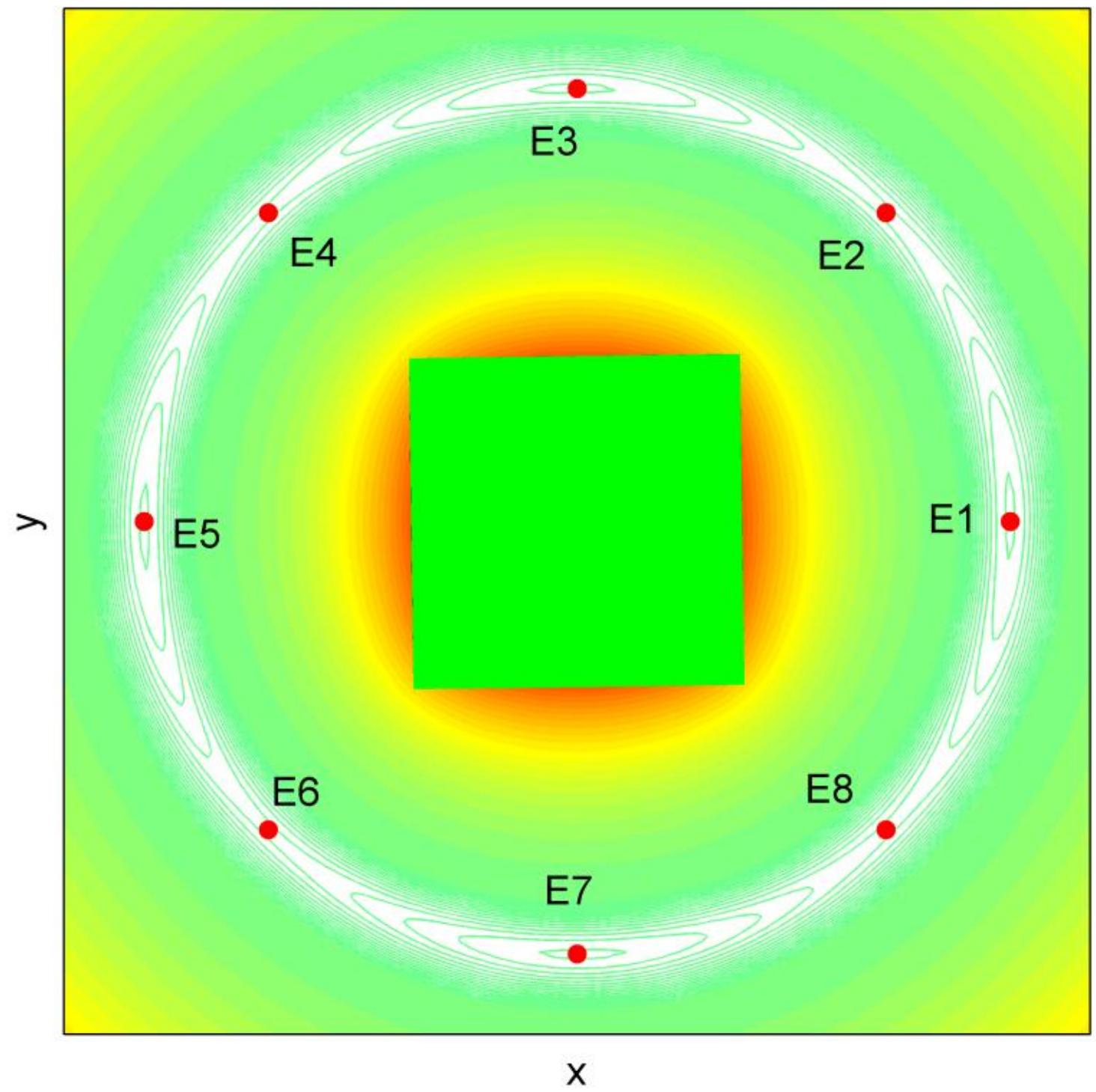

Fig. 4 The equilibrium points around a rotating homogeneous cube

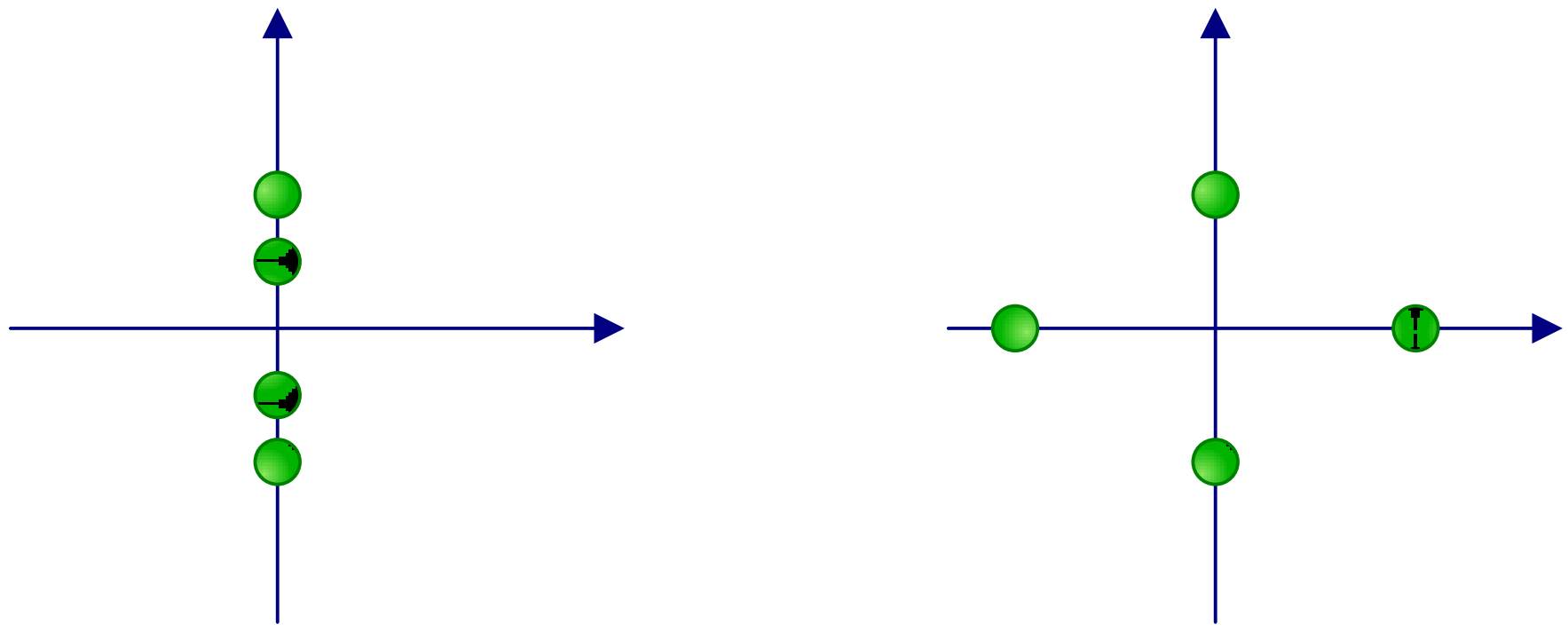

a. The distribution of the eigenvalues of the equilibrium points E1, E3, E5 and E7

b. The distribution of the eigenvalues of the equilibrium points E2, E4, E6 and E8

Fig. 5 The distribution of eigenvalues

Using the theory described in the previous sections, it is found that the distribution of the eigenvalues of the equilibrium points is that shown in Fig. 5. There are two families of periodic orbits in the xy plane near equilibrium points E1, E3, E5 and E7. There is only one family of periodic orbits in the xy plane near the equilibrium points E2, E4, E6 and E8. Thus, we obtain essentially identical results concerning the periodic orbits around the equilibrium points to those found by Liu et al. [33]. In addition, we can also draw further conclusions: There is only one family of quasi-periodic orbits on the xy plane near each of the equilibrium points E1, E3, E5 and E7, and these quasi-periodic orbits are on the 2-dimensional torus $T^2$.

**4.2 Application to Asteroid 1620 Geographos**

In this section, we apply the above contents to asteroid 1620 Geographos to compute the mutual annihilation of equilibrium points and the change of topological cases. For the equilibrium points E1, E2, E3, and E4, the ratio of the out-of-plane position and the maximum position in the xy plane are 0.033, 0.011, 0.051, and 0.013 [57]. Thus the relative value of the out-of-plane position is small, and the characteristic of plane-symmetric for asteroid 1620 Geographos is notable. 1620 Geographos has overall dimensions of $(5.0\times 2.0\times 2.1)\pm 0.15$ km [58], rotation period 5.222 h [59], and estimated bulk density $2.0\ \mathrm{g\cdot cm^{-3}}$ [58]. The gravitational field and geometrical shape of 1620 Geographos are computed by the polyhedral method [17] with data from [60]. The gravitation and shape of 1620 Geographos are calculated by 8192 vertices and 16380 faces. The YORP effect causes the rotation speed of asteroid 1620 Geographos magnify, the accelerated speed of the rotation equal $1.15\times 10^{-8}\mathrm{rad\cdot d^{-2}}$[61].

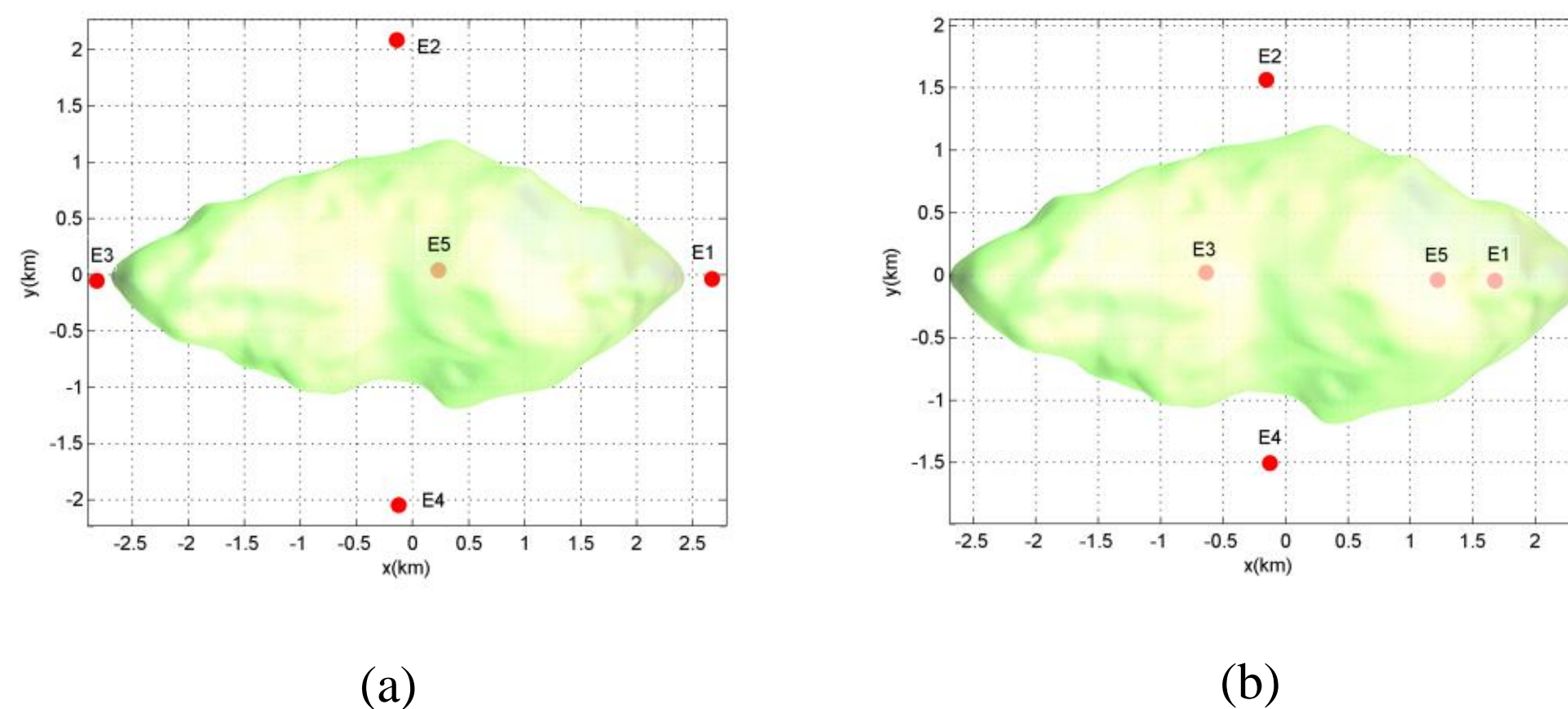

(a) (b)

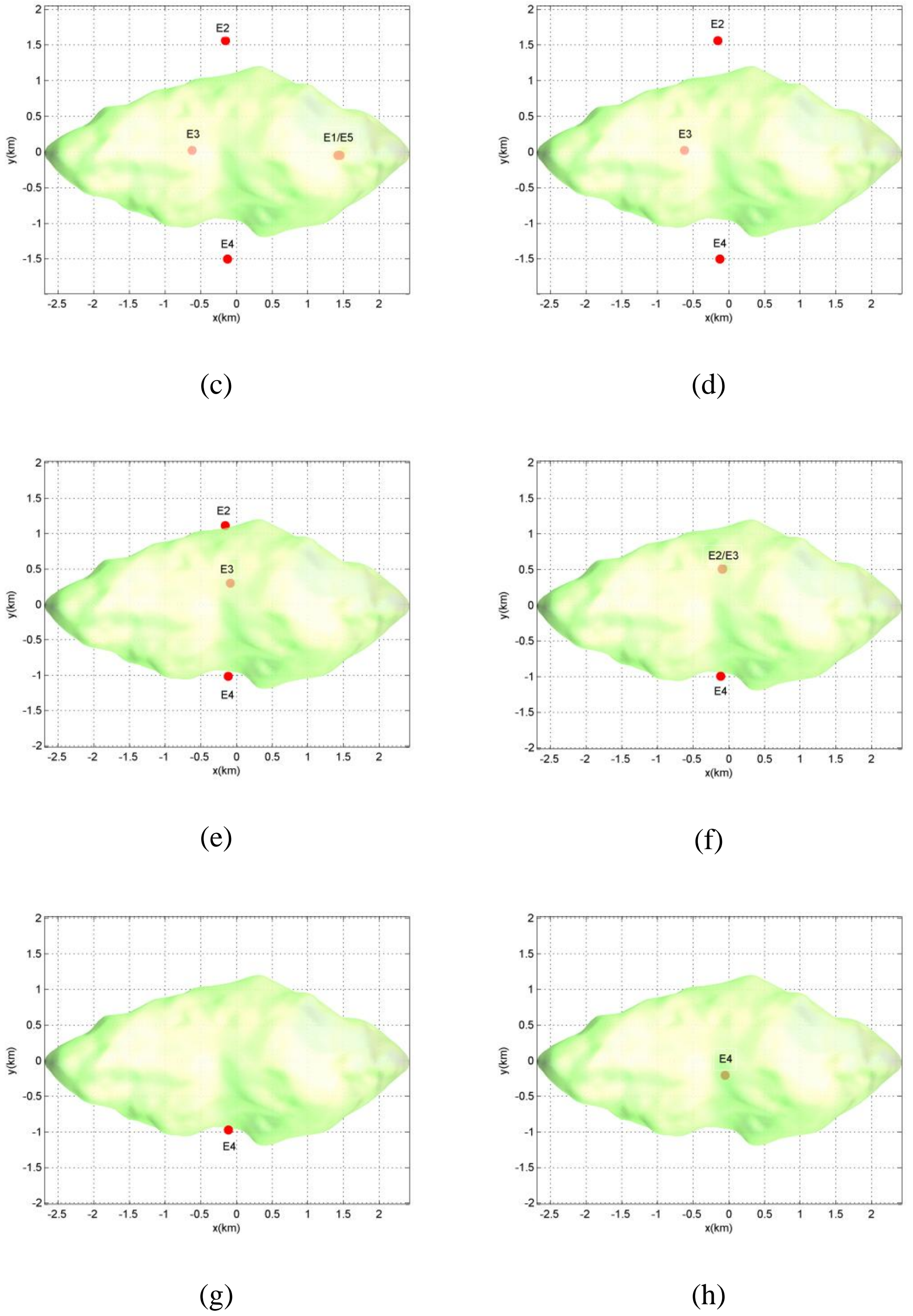

Fig. 6 The numbers and positions of relative equilibria of asteroid 1620 Geographos: (a) $\omega = \omega_0$; (b) $\omega = 1.47059\omega_0$; (c) $\omega = 1.47449\omega_0$; (d) $\omega = 1.47450\omega_0$; (e) $\omega = 2.2730\omega_0$; (f) $\omega = 2.32558\omega_0$; (g) $\omega = 2.32559\omega_0$; and (h) $\omega = 2.439\omega_0$.

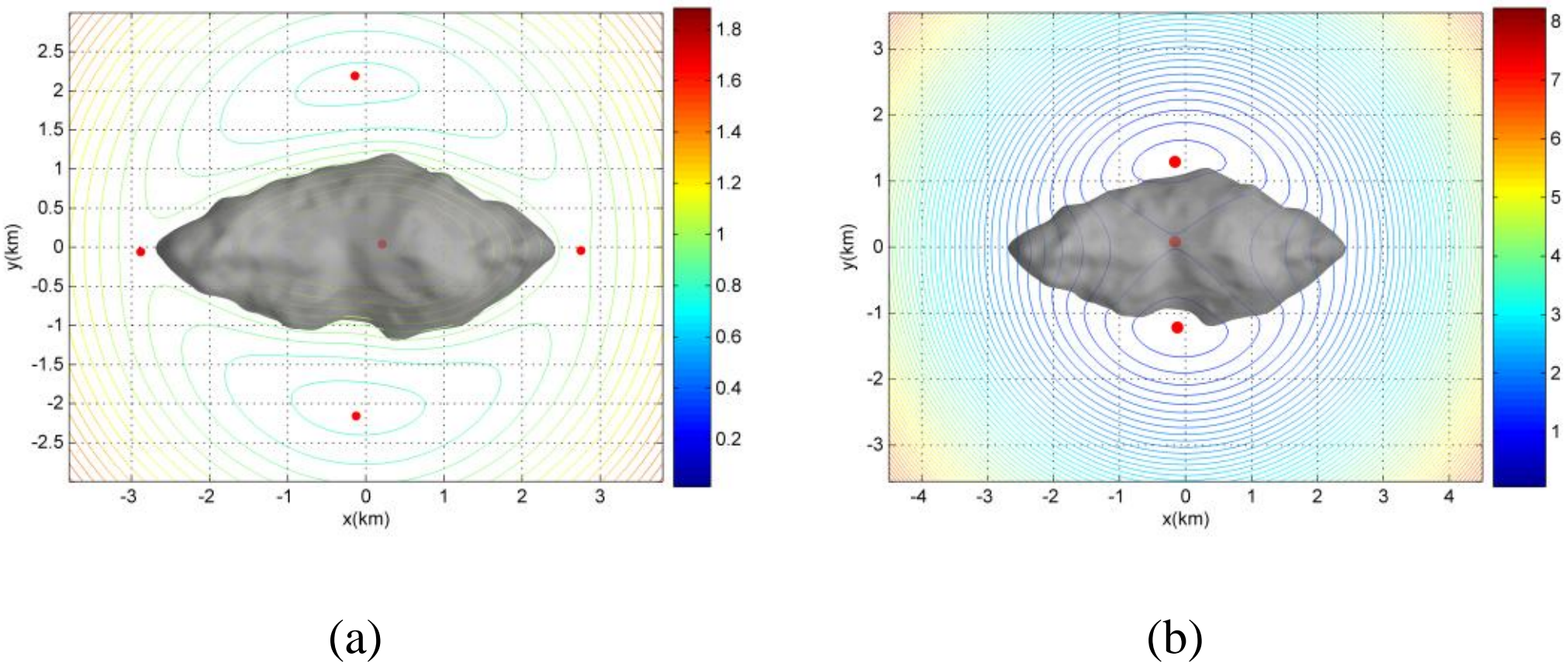

(a) (b)

Fig. 7 The contour plot of the zero-velocity surfaces of asteroid 1620 Geographos: (a) five equilibrium points with $\omega = \omega_0$; (b) three equilibrium points with $\omega = 2.0\omega_0$. The unit is $m^2s^{-2}$.

If the rotation speed changes, the number of equilibrium points changes. Fig. 6 shows the numbers and positions of relative equilibria of asteroid 1620 Geographos when the parameter changes. Fig. 7 shows the contour plots of the zero-velocity surfaces of asteroid 1620 Geographos when it has five and three equilibrium points, respectively. Bifurcations occur at Fig. 6c and Fig. 6f. Two equilibrium points collide and annihilate each other at $\omega = 1.47449\omega_0$ and $\omega = 2.32558\omega_0$. When $\omega = \omega_0$, there are five equilibrium points in the potential field of 1620 Geographos. If $\omega$ increases, equilibrium points E1 and E5 will approach each other; when $\omega = 1.47449\omega_0$, E1 and E5 collide. The topological cases of these two equilibrium points are different; E1 belongs to Case f while E5 belongs to Case a. The collided equilibrium point E1/E5 is degenerate. If $\omega$ continues to increase, there are only three equilibrium points left, they are E2, E3, and E4. E2 and E3 approach each other when $\omega$ increases in $[1.47450\omega_0, 2.32558\omega_0]$, and they collide and annihilate each other if and only of $\omega = 2.32558\omega_0$. E2 belongs to Case e while E3 belongs to Case f. When the mutual annihilation of equilibrium points at $\omega = 2.32558\omega_0$ occurs, the topological case of E4 becomes from Case e to Case k to Case a. This is the reverse process of the movement of eigenvalues showed in Fig. 2a. When $\omega = 2.32558\omega_0$, the topological case of E4 is Case k. Thus the Hopf bifurcation occurs in this situation. When $\omega > 2.32559\omega_0$, there is only one equilibrium point left, i.e. E4.

In addition, while the rotation speed changes, all the equilibria's positions change. When $\omega$ increases in $[1.0\omega_0, 1.47449\omega_0]$, E1 comes into the body of asteroid 1620 Geographos, and the position of mutual annihilation of E1 and E5 is in the body of asteroid 1620 Geographos. E3 also comes into the body of asteroid 1620 Geographos while E5 approach to E1. When E1 and E5 are annihilated, E3 is inside the body of asteroid 1620 Geographos, and E2 and E4 are outside. When $\omega$ increases in $[1.47450\omega_0, 2.32558\omega_0]$, E2 also comes into the body of asteroid 1620 Geographos, and the position of mutual annihilation of E2 and E3 is in the body of asteroid 1620 Geographos. After the mutual annihilation of E2 and E3, E4 is outside of the body of asteroid 1620 Geographos. When $\omega$ continues to increase, E4 will come into the body of asteroid 1620 Geographos.

To study the birth and evolution of the Solar System, the physical properties of the asteroids and comets, and the breakup of asteroids and the formation of binary asteroids are the scientific goals of several space missions [62-63]. If the rotation speed varies, the positions of the equilibrium points vary; surface shedding occurs when the equilibrium point touches the asteroid's surface [2]. From Fig. 6, one can see that the surface shedding begins near the intersection point of the –x axis and the surface during the changing process of the rotation speed. The rotation speed continues to change; the second surface shedding begins near the intersection point of the +x axis and the surface. From Fig. 6d, Fig. 6e, and Fig. 7b, one can see that if there are only three equilibrium points left, the surface shedding begins on each point of the asteroid 1620 Geographos' surface.

**5. Conclusions**

In this work, the dynamics of orbits near the equilibrium points in a rotating plane-symmetric potential field, including equilibrium points, periodic orbits, and manifolds, is studied. The structure of the submanifolds and subspaces near an equilibrium point was characterized. It has been found that there are twelve cases for the non-degenerate equilibrium points.

The necessary and sufficient conditions for linearly stable, non-resonant unstable and resonant equilibrium points are presented. A resonant equilibrium point is a Hopf bifurcation point. Periodic-orbit families are found to appear and disappear near resonant equilibrium points with parametric variation. Two non-degenerate equilibria may collide and change to one degenerate equilibrium point before mutual annihilation as the rotation speed changes.

The theory developed here is applied to the motion in the gravitational potential of a rotating homogeneous cube and asteroid 1620 Geographos. In the gravitational potential of a rotating homogeneous cube, it is found that there are two families of periodic orbits in the xy plane near equilibrium points E1, E3, E5 and E7, and there is only one family of periodic orbits in the xy plane near the equilibrium points E2, E4, E6 and E8. It has been found that the number and positions of equilibrium points in the potential of asteroid 1620 Geographos vary with the rotating speed of the asteroid.

## Acknowledgements

This research was supported by the National Basic Research Program of China (973 Program, 2012CB720000), the National Natural Science Foundation of China (No. 11772356, 11372150), and China Postdoctoral Science Foundation-General Program (No. 2017M610875).